%% file: paper.tex
\documentclass[sigconf]{acmart}

\AtBeginDocument{%
  }

\setcopyright{acmcopyright}
\copyrightyear{2018}
\acmYear{2018}
\acmDOI{XXXXXXX.XXXXXXX}

\acmConference[Conference acronym 'XX]{Make sure to enter the correct
  conference title from your rights confirmation emai}{June 03--05,
  2018}{Woodstock, NY}
\acmPrice{15.00}
\acmISBN{978-1-4503-XXXX-X/18/06}

\usepackage{epsfig}
\usepackage{hhline}
\usepackage{booktabs}
\usepackage{multirow}
\usepackage{subfigure}
\usepackage{autobreak}
\usepackage{tablefootnote}
\usepackage{amsmath,amsfonts} 
\usepackage{epsfig}
\usepackage{booktabs}
\usepackage{diagbox}
\usepackage{array}
\usepackage{multicol}
\usepackage{threeparttable}
\usepackage{epstopdf}
\usepackage{listings}
\usepackage{multirow}
\usepackage{subfigure}
\usepackage{makecell}

\newcommand{\figref}[1]{Figure \ref{#1}}

\newcommand{\tabref}[1]{Table \ref{#1}}
\newcommand{\secref}[1]{Section \ref{#1}}

\newcommand{\cut}[1]{}

\begin{document}

\title{Improving Topic Relevance Model by Mix-structured Summarization and LLM-based Data Augmentation}

\author{
	\begin{tabular}{ccccc}
		Yizhu Liu & Ran Tao & Shengyu Guo &  Yifan Yang & Zhenlin Wang \\
	\end{tabular}
}

\email{{liuyizhu, taoran04, shengyu.guo, yifan.yang}@meituan.com}
\affiliation{%
	\institution{Meituan Dianping}
	\city{Shanghai}
	\country{China}
}

\email{wangzhenlin@stu.scu.edu.cn}
\affiliation{%
	\institution{Sichuan University}
	\city{Chendu}
	\country{China}
}


\begin{abstract}
Topic relevance between query and document is a very important part of social search, which can evaluate the degree of matching between document and user's requirement. In most social search scenarios such as Dianping, modeling search relevance always faces two challenges. One is that many documents in social search are very long and have much redundant information. The other is that the training data for search relevance model is difficult to get, especially for multi-classification relevance model.
To tackle above two problems,
we first take query concatenated with the query-based summary and the document summary without query as the input of topic relevance model, which can help model learn the relevance degree between query and the core topic of document. Then, we utilize the language understanding and generation abilities of large language model (LLM) to rewrite and generate query from queries and documents in existing training data, which can construct new query-document pairs as training data.
Extensive offline experiments and
online A/B tests show that the proposed approaches effectively improve the performance of relevance modeling.
\end{abstract}




\maketitle

\input{intro}
\input{approach}
\input{eval}
\input{related} 
\input{conclude}

\bibliographystyle{ACM-Reference-Format}
\bibliography{relevance}


\end{document}

%% file: intro.tex
\section{Introduction}
\label{sec:intro}
In social search scenarios, topic relevance model aims at estimating the relevance between user query and document, which reflects the matching degree between document and user's requirements. In order to differentiate the degree to which user needs are satisfied, the topic relevance can be divided into 3 categories: \textit{strong relevance}, \textit{weak relevance} and \textit{irrelevance}. 
Strong relevance indicates that the document predominantly consists of information directly aligned with the query.
Weak relevance implies that the document contains only a limited amount of query-related information, like confined to a single sentence, with the remaining content being unrelated to the query.
Irrelevance signifies that the document's content is entirely unrelated or contradictory to the query.
The input of neural-based topic relevance model consists of query input and document input. 
Since the documents in social search are always very long and contain a variety of parts that can meet different queries, topic relevance model with whole document as input cannot effectively learn the alignment between query and document.
In addition, topic relevance models suffer from lacking reliable training data consisting of query and document pairs. Because data collected from online click behavior is not accurate enough, and human annotation is difficult and costly.
Thus, document input optimization and training data augmentation are two challenges of topic relevance modeling in social search scenarios.

\begin{table*}[ht]
	\centering
	\subtable[Examples of query-document pairs with relevance label in test set.]{    
		\begin{tabular}{|l|m{13.5cm}|m{2cm}<{\centering}|}
			\hline
			\textbf{Query} & \textbf{Document} & \textbf{Rel.} \\
			\hline
			\multirow{2}{*}{sakura} & \textbf{\em Doc 1:} \textbf{March is the perfect time to visit Yuyuantan Park in Beijing, a stunning spot to capture sakura. The best viewing time for sakura is usually in the middle to late March,} with only a week of full bloom that takes your breath away. Yuyuantan Park is particularly suitable for photography. \textbf{The combination of the TV tower and sakura creates incredibly beautiful photos.} & Strong \\
			\cline{2-3}
			 & \textbf{\em Doc 2:} Strolling through the hutongs in Beijing is an endlessly enjoyable activity, as it allows you to witness the ordinary lives of old Beijing while also experiencing a touch of artistic and cultural trends. The charming soul of these hutongs lies in the mix of taverns, restaurants, and small shops. From March to May, many flowers are in bloom, and \textbf{the sakura in Yuyuantan Park are particularly beautiful.}& Weak
			\\
			\hline
		\end{tabular}
	}
	\qquad
	\subtable[Optimized document input for \textit{Doc 2} from different approaches and relevance results predicted by relevance models trained with different inputs.]{   
			\begin{tabular}{|l|m{13.5cm}|m{2cm}<{\centering}|}
				\hline
				\textbf{Query} & \textbf{Optimized Document} & \textbf{Predicted Rel.} \\
				\hline
				\multirow{2}{*}{sakura} & \textbf{\em Query-focused:} From March to May, many flowers are in bloom, and \textbf{the sakura in Yuyuantan Park are particularly beautiful.} & Strong \\
				\cline{2-3}
				& \textbf{\em Mix-structured:} Strolling through the hutongs in Beijing is an endlessly enjoyable activity, \textit{(document summary)} \textbf{the sakura in Yuyuantan Park are particularly beautiful.} \textit{(query-focused summary)} & Weak
				\\
				\hline
			\end{tabular}
	}
	\caption{(a) shows the examples of strong relevance and weak relevance. (b) shows the performance of topic relevance models with document inputs of (a) optimized by different methods. The bold sentens are related to query. ``Rel.'' denotes relevance.}
	\label{tab:querydoc}  
\end{table*}

\begin{table}[h]
	\centering
	\begin{tabular}{|c|m{4.5cm}|m{1.2cm}|}
		\hline
		\textbf{Query} & \textbf{Document} & \bf Rel. \\
		\hline
	    Beef hot pot & My favorate Beef hot pot. I love her hot pot in mini form, the pot base is only 10 yuan, and the flavor of the pot base is delicious and spicy, very satisfying!  Overall, it's great. & Strong\\
		\hline
	\end{tabular}

	\begin{tabular}{|c|m{4.5cm}|m{1.2cm}|}
		\hline
	\textbf{Type} & \bf Query & \bf Rel. \\
		\hline
		Syn. rewrite & hot pot & Strong \\
		\hline
		Ant. rewrite & beef snacks & Irrelevant \\
		\hline
		Generation & {\bf mini hot pot}>pot base>spicy & Strong \\
		\hline
		Generation & mini hot pot>pot base>{\bf spicy} & Weak \\
		\hline
	\end{tabular}
	\caption{Examples of rewritten and generated queries with their corresponding documents. Queries from query generation are document keywords sorted according to their importance. Relevance label (Rel.) corresponds to the bold keyword.}
	\label{tab:llmquery}  
\end{table}

The common method of document input optimization is extracting query-focused summary from document as document input of relevance model~\cite{baidusearch21}. Such methods retain the coarse-grained relevant contents and discard the rest before measuring the semantic relevance between query and document, which can better assist the model in determining whether the document contains information that meets the user's needs. However, only using query-focused summaries as document input cannot teach the model to distinguish between strong relevance and weak relevance, because query-focused summaries cannot reflect the proportion of information in the document that is relevant to the query.
As shown in \tabref{tab:querydoc}, 
given one query ``sakura'', both \textit{Doc 1} and \textit{Doc 2} contain the information of sakura. 
The entire text of \textit{Doc 1} is describing sakura, representing strong relevance. \textit{Doc 2} only has one sentence describing the sakura, while other content describes the amusement facilities, representing weak relevance. 
But the similar query-focused summaries for \textit{Doc 1} and \textit{Doc 2} make the topic relevance model view them as strong relevance to query when using these summaries as modified document input. 
To solve above problem, we propose a \textbf{mix-structured summarization} method to improve document input of topic relevance model by connecting query-focused summary and document summary without considering query. As shown in \tabref{tab:querydoc} (b), mix-structured summaries refer to the information related to query in the document and the core information of the document, leading to a clear differentiation between strong and weak relevance.

As a neural-based model, topic relevance model requires sufficient data to train. The Existing method of collecting training data of topic relevance model involves manual labeling~\footnote{The data used to manually annotated is from the search logs of online search engines.} and click-through data filtering~\cite{yao2021learning}.
The former is costly, and the latter has a lot of noise. 
Meanwhile, it is rare to find completely irrelevant query and document pairs in real-world search engines, which leads to the trouble of introducing irrelevant training samples~\cite{jiang2019unified}.
Recently, Large Language Models (LLMs) excel at handling general instructions and have shown promise in data generation tasks~\cite{LLM2023selfinstruct,schneider-turchi-2023-team}.
We explore the methods for data augmentation in topic relevance model by using LLM. We leverage the capability of LLMs in Nature Langauge Understanding (NLU) and Nature Langauge Generation (NLG) to accomplish \textbf{query rewriting} and \textbf{query generation}. For query rewriting, we design prompts to obtain synonyms and antonyms of queries.
Synonyms (Syn.) have the same semantics as query but use different words, which will be paired with the corresponding document and labeled. Antonyms (Ant.) have the opposite semantics to query but use similar words. For query generating, we design a prompt to generate keywords from document and take extracted keywords as new queries. 
As shown in \tabref{tab:llmquery}, the rewritten and generated queries will be combined with the corresponding documents and labeled with reasonable relevance. 
LLM-based data augmentation can generate sufficient training samples for different relevance categories, which can help model training and enable the model to learn more knowledge.

In summary, this paper makes the following contributions:
\begin{enumerate}
\item We propose a simple but 
effective method, Mix-structured Summarization, to optimize the document input of topic relevance model. 
Mix-structured Summarization is composed of query-based summary and general summary of document, which can better differentiate the degree of relevance between query and document.
(\secref{sec:opt_input})

\item 
We explore the application of large language models in data augmentation for topic relevance model. Through LLM-based data augmentation, the topic relevance model can be better trained to learn alignment between query and document in different relevance categories.
(\secref{sec:llm_data})

\item 
 Offline and online experiments show that
the proposed approaches can improve the performance of relevance scoring in
social search. (\secref{sec:offline} and \secref{sec:online})
\end{enumerate}

Both mix-structured summarization and LLM-based data augmentation can be easily replicated and do not add to the burden of deploying models, making them well-suited for long document relevance tasks in real social search.

%% file: approach.tex
\section{Approach}
\label{sec:approach}
In this section, 
we first introduce our proposed mix-structured summarization method which combines query-focused summary and general summary as document input of topic relevance model, and then explain the LLM-based data augmentation in detail.

To specify the degree to which user needs are met, our goal is to categorize the relevance between queries and documents into three classes, i.e., strong relevance, weak relevance and irrelevance.
Strong relevance means that most of the information in the document is a description of the query.
Weak relevance denotes that the document contains only a small amount of information related to the query and the rest of the information is not relevant to the query.
Irrelevance represents that the information in the document is completely inconsistent with the query. 
We take a 3-class BERT model as our topic relevance model.

\subsection{Mix-structure Summarization} 
\label{sec:opt_input}
\begin{figure}[th]
	\centering
	\includegraphics[width=0.9\linewidth]{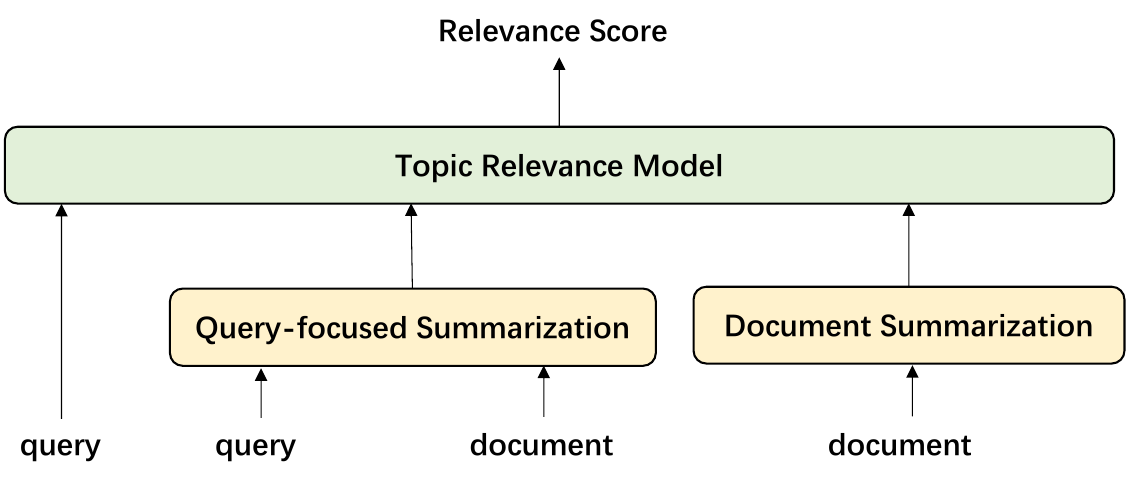}
	\caption{Illustration of mix-structured summarization.} 
\label{fig:optinput}
\end{figure}

Let $Q$ and $D$ denote the query and document separately.
To assess the relevance of query and document, 
it is necessary to consider both whether there is information related to the query in the document and the proportion of query-related information in the document. Thus, we first \textbf{extract query-focused summaries} and \textbf{common document summmaries} from $D$, and then connect them via the special token $SEP$ as mix-structured summary, as illustrated in \figref{fig:optinput}. Taking query and mix-structured summary as relevance model input can discern whether document is relevant and distinguish between strong relevance and weak relevance. 

\subsubsection{Query-focused  Summarization} 
Following \citet{baidusearch21}, we extract summary $S_{query}$ from $D$ with respect to a given query $Q$.
We tokenize $Q$ to obtain a set of query-based tokens $W_{q}$
We segment $D$ into sentences and tokenize each sentence. Each sentence $s$ has a set of tokenized words $W_{s}$. To cover different tokens in $W_{q}$,
each token of $W_{q}$ will be signed a sentence,  the $W_{s}$ of which includes the token.         
Besides, 
we sequentially combine the selected sentences with their adjacent sentences in the document, until the length of the merged sequence meets the predetermined threshold. In this way, we can get a query-focused summary, expressing query related information and its contextual information in the document.

\subsubsection{Document  Summarization} 
In order to distinguish between strong and weak relevance documents that both contain query related information, we extract common summary $S_{doc}$ from $D$ without considering query.
$S_{doc}$ should be comprehensive, capable of representing all the key points in $D$. 
Inspired by Lead-3~\footnote{Lead-3~\cite{www2023} is a commonly-used extractive summarization baseline that selects first three sentences of document as summary. Although it is the simplest method, it can yield a fairly good result.}, 
we select the first 3 sentences of each paragraph in document and concatenate all selected sentences as summary.  Such extractive summary comprehensively abstracts the main information of the document.


\subsection{LLM-based Data Augmentation} 
\label{sec:llm_data}
It is difficult and costly to obtain sufficient and correct training data for topic relevance model. The training data collected from real-world search engines lacks weak relevance and irrelevance data. Therefore, we explore the use of LLM to augment training data. 

Each sample of topic relevance training data consists of three parts: \textit{query}, \textit{document} and \textit{label}. 
As shown in \figref{fig:llm}, we expend relevance training data of different categories by \textbf{query rewriting} and \textbf{query generation}.

\subsubsection{Query Rewriting} 
We design two prompts (as shown in \figref{fig:llm_rewrite}) to generate synonyms and antonyms of queries in the existing dataset.
To increase the complexity of the synthetic data, our prompts contain some constraints for generation.
We also provide some specific examples in the prompts, which help LLM understand our requirements better.

\begin{figure}[th]
	\centering
	\includegraphics[width=1\linewidth]{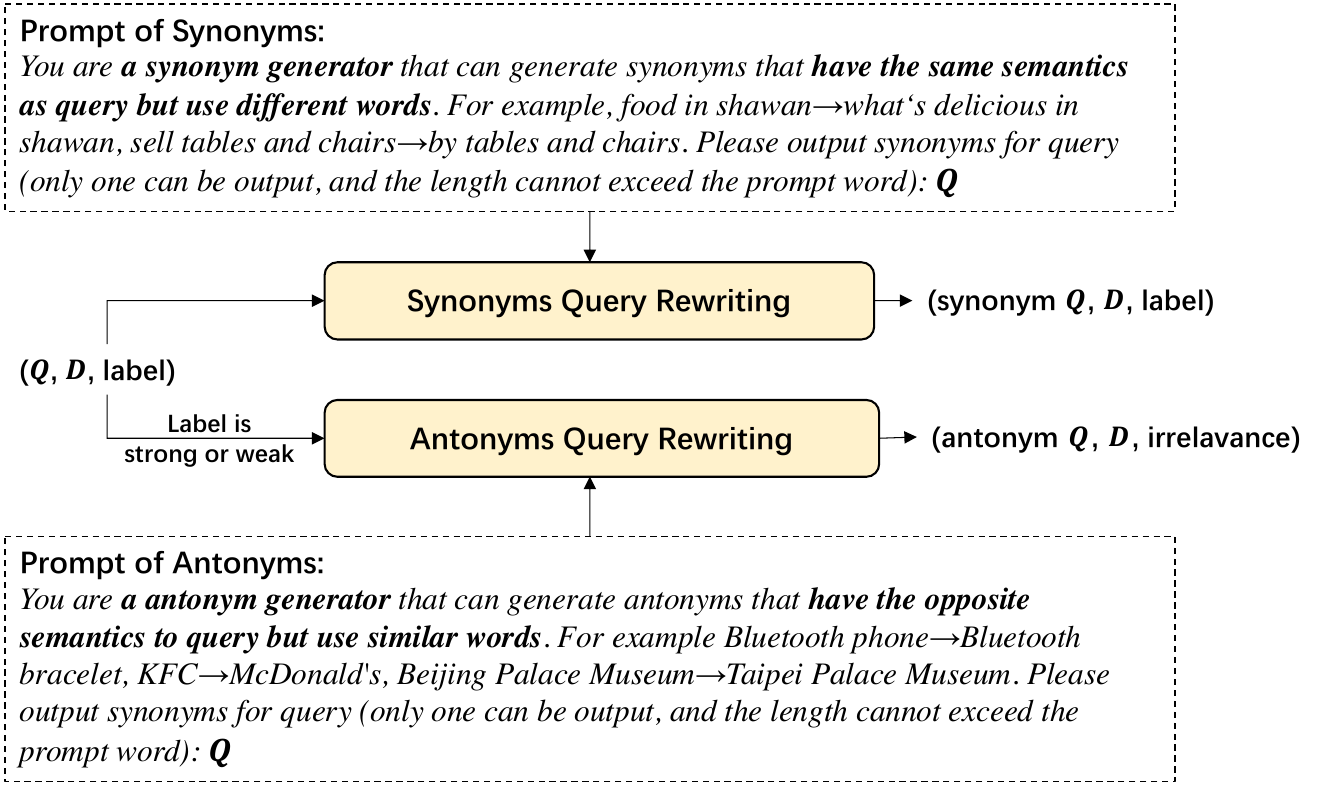}
	\caption{The process of LLM-based query rewriting. Tuples of (Q, D, label) are new samples.} 
\label{fig:llm_rewrite}
\end{figure}

For synonyms query rewriting, given a query $Q$, the prompt informs the LLM that it is a synonym generator and asks it to generate synonyms for $Q$.  The synonyms of $Q$, together with the corresponding document of $Q$, constitute new training samples. Their labels align with the label of $Q$ and its associated document.
This method is able to construct relevant data across various categories.
Furthermore, in order to enhance the robustness of the training set, the prompt also limits LLM to generate synonyms that \textit{have the same semantics as query but use different words}
and shows some hard-to-distinguish synonyms. 
We expect LLM to generate high-quality and challenging training sets, which can assist in improving the robustness of relevance model.

For antonyms query rewriting, we select strong relevance and weak relevance training data. Similar to synonyms query rewriting, LLM is asked as an antonym generator, which needs to generate antonyms that \textit{have the opposite semantics to query but use similar words}. The antonyms of query $Q$, together with the corresponding document of $Q$, should be labeled as irrelevance.

\subsubsection{Query Generation} 
Document in strong relevance and weak relevance data must contain query related information. Given a document $D$, we ask LLM to extract some keywords from the document, which can be combined with the given document as strong relevance or weak relevance samples.
In the keyword generation prompt, we require LLM to extract 3 keywords and output them in order of their importance in the document.
Keywords with the most importance and the given document form a strong relevance sample.
Keywords with the lowest importance and the given document form a weak relevance sample.

\begin{figure}[th]
	\centering
	\includegraphics[width=1.0\linewidth]{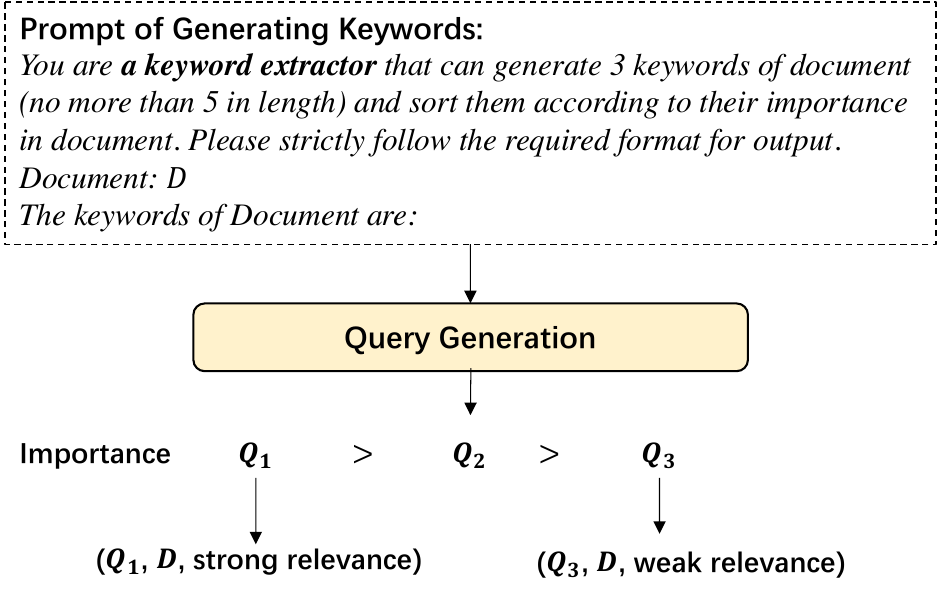}
	\caption{The process of LLM-based query generation. Tuples of (Q, D, label) are new samples.} 
\label{fig:llm_rewrite}
\end{figure}

By incorporating above new training samples, we can effectively enhance the diversity and complexity of the training set, thereby improving the model's ability to handle a wider range of language scenarios.

%% file: eval.tex
\section{Evaluation}
\label{sec:eval}
To demonstrate the effectiveness of our proposed method, we conduct experiments on an industrial social search engine. 
Experimental results show that for social search with long documents, the proposed mix-structured summarization and LLM-based data augmentation can significantly improve the performance of topic relevance modeling.

\subsection{Datasets}
In this experiment, 
the relevance dataset consists of 
human-labeled query-document pairs from historical search results from Dianping Social Search. 
Training set contains 46,306 samples, validation set contains 2,500 samples, and test set contains 8514 samples.
With the help of Dianping's crowd-sourcing platform, each of these query-document pairs is annotated as one of the three relevance categories: strong relevance, weak relevance, and irrelevant. 
Specifically, there are 33,453 strong
relevant samples, 6,530 weak relevant samples, and 6,323 irrelevant samples in training set.

\subsection{Experimental Setup}
We take BERT-based cross-encoder~\cite{bert2019} as basic model, as it is widely used for relevance tasks in industrial search engines. 
The methods compared in this experiment are listed below.

\begin{itemize}
	\item \textbf{Baseline} is a cross-encoder model with query-focused summary as document input. \citet{baidusearch21} showed that relevance modeling can be benefitted by query-focused summary.
	
	\item \textbf{MSD-CE} is a cross-encoder model with mix-structured summary as document input. 
	
	\item \textbf{LLM-CE} is a cross-encoder model with query-focused summary as document input, which is trained on data augmented by query rewriting and query generation.
	
	\item \textbf{MSD-LLM-CE} is a cross-encoder model with mix-structured summary as document input, which is trained on data augmented by query rewriting and query generation.
\end{itemize}

To be fair, the max input length of all relevance models is set as $192$. For models with mix-structured summaries as document input, the max length of query-focused summaries is $128$ and the max length of comment document summaries is $64$. LLM used in our experiments is GPT-3.5-turbo.

\subsection{Evaluation Metrics}
We use the following evaluation metrics in this experiment.

\textbf{Area Under Curve (AUC) }~\cite{jiang2019unified, yao2021learning}
is widely used for evaluating relevance models. It is an offline metric, which is used on test set.
As our relevance task has three categories, Inspired by~\citet{zan2023spm}, we compute multiclass AUC in our experiments as follows:
\begin{equation}
	f(a, b)=\left\{
	\begin{array}{rcl}
		1 & & {a > b}\\
		0 & & {others}\\
	\end{array} \right.
\end{equation}

\begin{equation}
	AUC = \frac{\sum_{j}\sum_{k} f(y_j, y_k) \cdot f(p_j, p_k)}{\sum_{j}\sum_{k} f(y_j, y_k)}
\end{equation}

where $y_j$ and  $p_j$ are the reference relevant score and predicted relevant score for the $j^{th}$ query and document pair respectively.
The Relevant score for strong relevance, weak relevance, and irrelevant are 1, 0.7, and 0.

\textbf{Good vs. Same vs. Bad (GSB)}~\cite{zhao2011automatically} is a professional human evaluation metric, which
compares the search results from two same search engines with different topic relevance models.
Given a user query and two returned documents from different relevance models, the annotators, who do not know which relevance model the result belongs to, are required to label the result with better relevance as ``Good'' and the one with poor relevance as ``Bad''. If these two results are equally good or bad, both of them should be labeled as ``Same''. 
As the query and return document pairs evaluated by GSB are obtained from online sources,
Following \citet{yao2021learning}, we use $\Delta GSB$ to reflect real user satisfaction.
\begin{equation}
	\Delta GSB= \frac{\#Good - \#Bad}{\#Good + \#Same + \#Bad} 
\end{equation}

\subsection{Offline Experimental Results}
\label{sec:offline}
In this section, we analyze the performance of our proposed methods on an offline test set. 
As shown in \tabref{tab:result_offline}, after taking mix-structured summaries as document input and augmenting training data by LLM, MSD-LLM-CE outperforms all other models.

\begin{table}[th]
	\centering
	\begin{tabular}{|l|c|}
		\hline
		 \bf Model & \bf AUC (\%) \\
		\hline
		 Baseline & 73.64 \\
		 MSD-CE& 75.71 \\
		 LLM-CE& 75.94 \\
		 MSD-LLM-CE& \bf 77.41\\
		\hline
	\end{tabular}
	\caption{Performance of different topic relevance models.}
	\label{tab:result_offline}
\end{table}

\subsubsection{Query-focused Summarization vs. Mix-structured Summarization} 
Baseline and LLM-CE only take query-focused summaries as document input, which ensures that the model can identify information in the document that matches the query.
MSD-CE and MSD-LLM-CE take mix-structured summaries as document input, 
which provides more comprehensive information of document.
As shown in \tabref{tab:result_offline}, MSD-CE and MSD-LLM-CE perform better than Baseline and LLM-CE respectively, demonstrating
that mix-structured summaries can assist the model in identifying query-related information in the document and determining the proportion of query-related information in the document. 
\tabref{tab:querydoc} shows that query-focused summary may cause the model to confuse strong relevance and weak relevance. Compared with query-focused summaries, mix-structured summaries can guide the model to better distinguish between strong relevance and weak relevance.

\subsubsection{With LLM-based Data Augmentation vs. Without LLM-based Data Augmentation}
As shown in \tabref{tab:augdata}, after applying LLM-based data augmentation method on training dataset, 
the total number of training samples has tripled. 
In particular, the quantity of weak relevance and irrelevance training samples is significantly increased.
In \tabref{tab:result_offline}, LLM-CE and MSD-LLM-CE outperform Baseline and MSD-CE respectively, 
indicating that it is feasible to use LLM for data augmentation. 
LLM-based data augmentation enhances the quantity and diversity of training data, 
which helps improve the generalization ability and robustness of the model.

\begin{table}[th]
	\centering
	\begin{tabular}{|l|c|c|c|c|}
		\hline
		\bf Training set & \bf Strong & \bf Weak & \bf Irrelevance & \bf Overall \\
		\hline
		w/o Augmented & 33,453 & 6,530& 6,323 & 46,306 \\
		\hline
		with Augmented & 41,946 & 39,564 & 42,386 & 123,896 \\
		\hline
	\end{tabular}
	\caption{Training data statistics for different relevance categories.}
	\label{tab:augdata}
\end{table}

\subsection{Online Experimental Results}
\label{sec:online}
Online A/B test between old online relevance model and new online relevance model using our proposed approach shows that the new model can improve the overall user experience in social search scenarios. Moreover, randomly sampling 100 user queries, $\Delta GSB$ between old online relevance model and new online relevance model is 0.1, which means new online model benefits from our proposed approach.

%% file: related.tex
\section{Related Work}
\label{sec:related}
In social search scenarios, the query and document (query-doc) are very different in length, and the
query may match only part of the document. 
Neural methods~\cite{dai2018convolutional, huang2013learning, mitra2017learning, wu2017sequential, bert2019, zan2023spm} have been flourishing for their better capability to model semantic similarity.

Typical query-document search relevance modeling~\cite{xu2018deep, guo2016deep, xiong2017end, dai2018convolutional, baidusearch21} can be divided into two classes:
global distribution of matching signals and local distribution of matching signal.
Models based on global distribution first calculate the matching signals among the document
for each query and then aggregate the distribution. 
\citet{guo2016deep} summarized the word-level similarities by pyramid pooling. 
\citet{xiong2017end} introduced kernel-pooling
to summarize the translation matrix and provide soft match signals.
\citet{dai2018convolutional} used CNN to represent n-grams of different lengths and softly match them in a unified embedding space.
As for models based on local distribution, they first extract the local context
of each query among document and conduct matching between
query and local context, and then aggregate the local matching
signals for relevance.
\citet{pang2017deeprank} proposed DeepRank to extract relevant contexts and
determine relevance score of each context first and then aggregate the distribution of all local contexts.
\citet{baidusearch21} extracted local relevance context of query from document and input the concatenation of query and local context to the BERT model to compute their relevance. 
Above two typical query-doc relevance models are not good at multi-class relevance tasks, becase they cannot effectively balance the recognition of query related information and the proportion of query related information.
Therefore, we propose to generate a mix-structure summary consisting of query-focused summary and common-document summary as document input of relevance model.

In recent years, BERT-based models~\cite{bert2019} get great success in relevance tasks.
However, such BERT-based models demand large-scale training data.
Collecting data by human annotation is costly and inefficient. 
Annotated query and document according to user click is not accurate enough, as user click is not only affected by relevance. 
It is common to encounter an imbalanced distribution of training data across different relevance categories, since
completely irrelevant query and document pairs are difficult to be found in real-world search engines~\cite{jiang2019unified}.
Data augmentation is an intuitive method to deal with above problems.
Recent research shows that LLMs demonstrate excellent
generalization performance across various natural language understanding and natural language generation tasks in few-shot or zero-shot scenarios~\cite{openai2023gpt4}. 
Thus, we explore generating query-doc training samples through LLM-based data augmentation, such as query rewriting and query generation.

%% file: conclude.tex
\section{Conclusion}
\label{sec:conclude}
In this work, we proposed two methods to enhance the multi-classification topic relevance model: 1) took query-based summary and document summary without query as document input of the topic relevance model, and 2) rewrite and generate query from query and document separately by large language model to construct new training data.
The results showed that our methods can make full use of document information to learn the relevance between query and document and improve topic relevance model through data augmentation.